\documentclass[11pt]{article}
\pdfoutput=1
\usepackage{graphicx}
\usepackage{amsmath,amsthm, amssymb}
\usepackage{graphicx}
\usepackage{hyperref}
\usepackage{booktabs}
\newcommand{\be}{\begin{eqnarray}}
\newcommand{\ee}{\end{eqnarray}}

\def\half{\frac{1}{2}}
\def\nn{\nonumber}

\begin{document}
\thispagestyle{empty}

 \renewcommand{\thefootnote}{\fnsymbol{footnote}}
\begin{flushright}
 \begin{tabular}{l}
 {\tt arXiv:1205.0069[hep-th]}\\
 {KIAS-P12023}
 \end{tabular}
\end{flushright}

 \vfill
 \begin{center}
 {\bfseries \Large Superconformal Index with Duality Domain Wall}
\vskip 1.9 truecm

\noindent{{\large
  Dongmin Gang, Eunkyung Koh 
and Kimyeong Lee \footnote{arima275, ekoh, klee(at)kias.re.kr} }}
\bigskip
 \vskip .9 truecm
\centerline{\it Korea Institute for Advanced Study, 
Seoul 130-012, Korea}
\vskip .4 truecm
\end{center}
 \vfill
\vskip 0.5 truecm

\begin{abstract}
We study a superconformal index  for ${\cal N}=4$ super Yang-Mills on $S^1 \times S^3$ with a half BPS duality domain wall inserted at the great two-sphere in $S^3$. 
The index is obtained by coupling the 3d generalized superconformal index on the duality domain wall with  4d  half-indices.  We further consider insertions of line operators to the configuration and propose integral equations which express that the 3d index on duality domain wall  is a duality kernel relating half indices of two line operators related by the duality map.  
We explicitly check  the proposed integral equations for various duality domain walls and line operators in the ${\cal N}=4$ $SU(2)$ theory. We also  briefly comment on a generalization to  $\mathcal{N}=2$ $A_1$ Gaiotto theories with a simple example, ${\cal N}=2$ $SU(2)$ SYM with four flavors.

\end{abstract}

\vfill
\vskip 0.5 truecm

\setcounter{footnote}{0}
\renewcommand{\thefootnote}{\arabic{footnote}}

\newpage

\tableofcontents

\section{Introduction and Conclusion}

Exact calculations in supersymmetric theories
are remarkable achievements in recent years. One early example is the partition function for 4d ${\cal N}=2$ gauge theories in the Omega-background \cite{Nekrasov:2002qd}. 
Recently, it is extended to the $S^4$ partition function with insertions of line operators  \cite{Pestun:2007rz, Gomis:2011pf}, $S^3$ partition function for 3d ${\cal N}=2$ theories \cite{Kapustin:2009kz,Jafferis:2010un, Hama:2010av} and the $S^4$ partition function with insertions of certain supersymmetric domain walls \cite{Drukker:2010jp, Hosomichi:2010vh}.  The 4d superconformal index \cite{Romelsberger:2005eg, Kinney:2005ej}, a twisted partition function on $S^1\times S^3$, is another exactly calculable quantity. The calculation is extended for 3-dimensional gauge theories on $S^1 \times S^2$ \cite{Kim:2009wb, Imamura:2011su} where the magnetic monopole operators are taken into account.  The 4d superconformal index with insertions of non-local operators is also considered. For instance, the line operator index is obtained in \cite{ Dimofte:2011py, Gang:2012yr} and the  surface operator index is studied in \cite{Nakayama:2011pa}.
 
 In this paper, we study the index in the presence of both 1/2 BPS duality domain walls and 1/2 BPS line operators in ${\cal N}=4$ super Yang-Mills theory.
Our approach employs two powerful tools developed recently. The first one is the generalized superconformal index for 3 dimensional theories \cite{Kapustin:2011jm}.
The generalized index is  defined by not only introducing the chemical potential for global symmetries
but also turning on magnetic flux of  the fictitious  gauge field for the global symmetries. 
The second one is the superconformal half-index for 4-dimensional theories which was introduced in  \cite{Dimofte:2011py} and also used profitably in studies of the index with line operators  in \cite{Dimofte:2011py, Gang:2012yr}.
The half-index is to a superconformal  index as  the Nekrasov partition function to a $S^4$ partition function.   As we  obtain the $S^4$ partition function by gluing two Nekrasov partition functions, the index on $S^1 \times S^3$ can be obtained by gluing two half-indices. As for the $S^4$ partition function with 3d domain walls,  the 4d superconformal index with 3d domain walls can be obtained by coupling the 3d generalized index on the domain wall with 4d half-indices.

 Duality domain walls in 4d $\mathcal{N}=4$ theories are first introduced in \cite{Gaiotto:2008sa,Gaiotto:2008ak} in the study of $S$-duality of supersymmetric boundary conditions. Recently, the duality domain walls draw much attention  as  a useful laboratory for studies of the 3d-3d version of AGT relation \cite{Alday:2009aq}. See \cite{Terashima:2011qi, Terashima:2011xe, Dimofte:2011ju, Dimofte:2011py} and references therein. 
 Field contents and Lagrangian for 3d theories on some duality domain walls in 4d ${\cal N}=4$ SYM are known \cite{Gaiotto:2008ak,  Terashima:2011qi}. However, such information for 3d theories on duality domain walls in general 4d ${\cal N}=2$ theories is limited in literatures.  In  \cite{Hosomichi:2010vh}, it is shown that the S-duality kernel in Liouville theory on a torus can be mapped to the $S^3$ partition function for $T[SU(2)]$ theory which is the theory on the S-duality domain wall in 4d ${\cal N}=4$ SYM with gauge group $SU(2)$. Using the same map, the $S^3$ partition function on the duality domain wall in 4d ${\cal N}=2$ $SU(2)$ SYM with four flavors is obtained from the duality kernel in Liouville theory on a sphere with 4 punctures  \cite{Teschner:2012em}.  Some speculations on the interpretation of the partition function as that of 3d gauge theory are also given in \cite{Teschner:2012em}. However, the theory on the duality domain wall is not yet identified to the authors' knowledge.

The organization of this paper is as follows.  In section 2, we review the construction of the duality domain wall, and the superconformal index compatible with the BPS duality domain walls and line operators.
In section 3,  we start with a general discussion on a relation between the generalized index on the duality domain wall and the half-index  on the hemi-sphere. 
Then we discuss how to add line operators into our setting. It leads to a proposal of integral equations for the index
on the duality domain walls and the half-index on the hemisphere in section 3.2.  This is inspired by the similar equation for the $S^3$ partition function for the self-dual wall and the $S^4$ partition function for 4d ${\cal N}=4$ SYM theory \cite{Hosomichi:2010vh}. 
To be concrete, in section 4,  we focus on the self-dual walls and line operators in the  ${\cal N}=4$ $SU(2)$  SYM theory. We show that the 3d index on the self-dual domain wall satisfies the proposed integral equations with 4d half-index with line operators. In section 5, we consider a generalization of the previous argument to the duality domain walls in generic ${\cal N}=2$ Gaiotto theories \cite{Gaiotto:2009we} of type $A_1$. We use the integral equations to obtain a few lowest orders of the index for the theory on the S-duality domain wall in 4d ${\cal N}=2$ $ SU(2)$ with four flavors theory. 
We confirm that the index coincides with the index for 3d theory proposed in  \cite{Teschner:2012em}.  We have two appendices.  We consider the 3d index of $T[SU(2), S^2]$ theory in Appendix \ref{app:1}.
The index of $T[SU(3)]$ theory and S-duality of the fundamental Wilson line in 4d $SU(3)$ ${\cal N}=4$ theory are discussed in Appendix \ref{app:2}. The result suggests that
acting S-duality twice maps the fundamental Wilson line to the anti-fundamental Wilson line. 

\section{Reviews}

In this section, we will review some background materials relevant to this paper. 

\subsection{A Duality Domain Wall} \label{duality wall}

Let us first briefly review the construction of a S-duality domain wall  in ${\cal N}=4$ SYM on ${\bf R}^{4}$ \cite{Gaiotto:2008sa,Gaiotto:2008ak}.

\begin{figure}[h!]    
\begin{center}
   \includegraphics[width=0.9\textwidth]{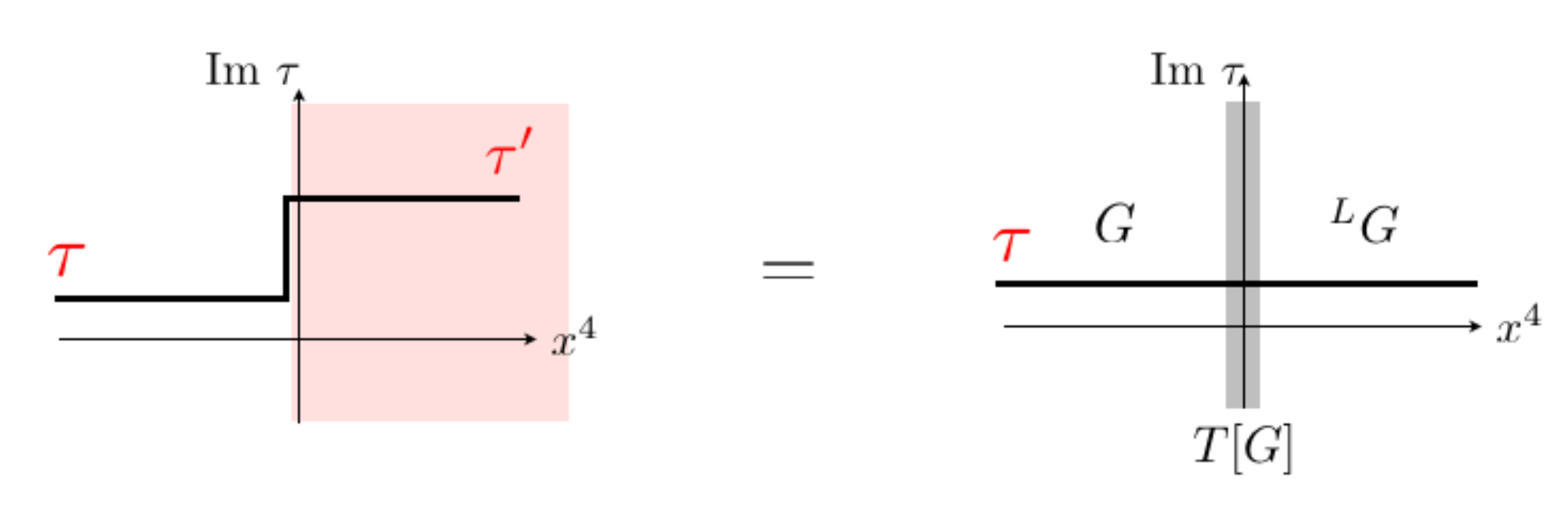} 
   \end{center}
   \caption{A Janus domain wall of 4d, ${\cal N}=4$ SYM with gauge group $G$ is a configuration that the holomorphic coupling $ \tau$ varies as a step function at $x^4=0$ (left). For $ \tau^{ \prime}$ related with $ \tau$ by S-duality, one can take S-duality on the right-half plane (red region), which results in the S-duality domain wall supported at $x^4=0$ (right).}
    \label{fig:domain}
\end{figure}
Consider a Janus domain wall of ${\cal N}=4$ SYM which is a configuration that the holomorphic coupling $ \tau := \frac{i 4 \pi}{g^2_{YM}} + \frac{ \theta}{ 2 \pi}$ varies as a step function of one of spatial directions, say $x^4$ (see the left side of fig.~\ref{fig:domain}) \cite{Bak:2007jm,  Gaiotto:2008sd}.  The holomorphic coupling is $ \tau$ on the right half-plane $x^4 < 0$ and $ \tau^{ \prime}$ on $x^4>0$. The Janus domain wall can preserve half of supersymmetries without additional degrees of freedom by breaking the SO(6) R-symmetry to SO(3) $ \times$ SO(3) \cite{Clark:2004sb, D'Hoker:2006uv, Kim:2008dj}
\footnote{A more general Janus configuration (holomorphic coupling as a general function $\tau(x^3)$) has been considered and shown to preserve half of supersymmetries. When $ \tau(x^3)$ is a step function,  conformal symmetry can be preserved thus we can use the conformal map from ${\bf R}^4$ to $S^1 \times S^3$ or $S^4$. 
}
. 
Taking S-duality on one of the half-planes, say $x^4>0$, the system becomes 4d ${\cal N}=4$ SYM with gauge group $G$ on the left half-plane and the same theory with ${}^L G$ (the Langlands dual or GNO dual of $G$) on the right half-plane. 
Also additional degrees of freedom appear at $x^4=0$ which can be described as a domain wall at $x^4=0$. Especially when  $\tau^{ \prime}$ is related with $ \tau$ by S-duality, it is called an S-duality domain wall. If $ \tau^{ \prime}$ is related with $ \tau$ by a general element $ \varphi$ in $PSL(2,{\mathbb Z})$ duality, it is called a duality domain wall.

\begin{figure}[h!]
\begin{center}
   \includegraphics[width=1\textwidth]{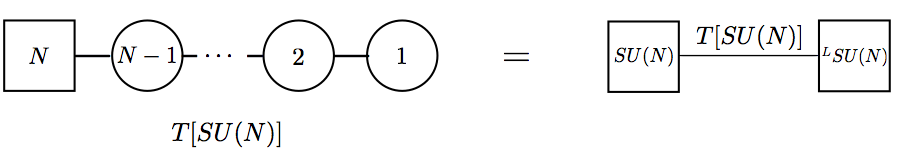} 
   \end{center}
   \caption{The quiver diagram of $T[SU(N)]$ theory (left). A unitary gauge group is denoted by a circle, $SU(N)$ flavor symmetry by a square, and a bi-fundamental hypermultiplet by a line.  $T[SU(N)]$ theory has a global symmetry $SU(N)\times {}^L SU(N)$ and we draw the quiver diagram for the theory  in a simpler form (right).   }
      \label{fig:quiver}
\end{figure}

\begin{figure}[h!]
\begin{center}
   \includegraphics[width=.5\textwidth]{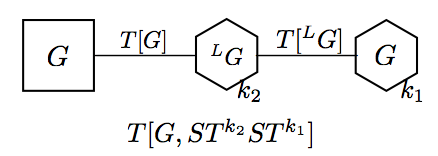} 
   \end{center}
   \caption{The quiver diagram for $T[G, \varphi]$ with $\varphi =S T^{k_1}  S T^{k_2} $. An integer $k$  under a hexagon  denotes the Chern-Simons level for  the gauge field (or background gauge field coupled to global symmetry) in the hexagon.  An internal node denotes a   gauge symmetry and external two nodes denote global symmetries.  }
      \label{fig:quiver2}
\end{figure}

For the S-duality domain wall, let us take the limit that ${\rm Im}[ \tau]$ is very large, then 4d theories are decoupled from the degrees of freedom on the domain wall.
 The remaining 3d theory is called $T[G]$ \cite{Gaiotto:2008ak}. If one starts with a Janus domain wall with $ \tau^{ \prime} = \varphi ( \tau)$, the resultant 3d theory can be denoted as $T[G, \varphi]$, adapting the notation in \cite{Terashima:2011qi}. In this notation, $T[G]$ can be understood as $T[G,\varphi=S]$. Explicit quiver diagrams for $T[G]$ for $G=SU(N), SO(2N+1), Sp(2N), SO(2N)$ are given in \cite{Gaiotto:2008ak}. Quiver diagrams for $T[SU(2), \varphi]$ are given in \cite{Terashima:2011qi}. We reproduce some of them in fig~\ref{fig:quiver} and \ref{fig:quiver2} which are relevant to our discussion. 

  $T[G, \varphi]$ theory has a global symmetry $G \times {}^L G$. The Higgs branch has  a classical symmetry $G$, manifestly rotating hypermultiplets, while the Coulomb branch has symmetry ${}^L G$  in the infrared. The symmetry enhancement at IR is due to monopole operators of 3d \cite{Gaiotto:2008ak}. If the ${\rm Im}[ \tau]$ is finite, the global symmetry $G$ (resp. ${}^L G$) is gauged with the 4-dimensional gauge group on the left (resp. right) half-plane to make up a 3d/4d coupled system. 

\subsection{4d and 3d Superconformal Indices}

In this section, we will consider 4d and 3d superconformal indices compatible with the duality domain wall.
 The half BPS domain wall ${\bf R}^3 \subset {\bf R}^4$
can be conformally mapped to $S^1 \times S^2 \subset S^1 \times S^3$. Let coordinates of ${\bf R}^4$ be $(\vec{x}, x^4)$ with $\vec{x}:=(x^1, x^2, x^3)$, where the conformal map is given by\footnote{Under the conformal map, ${\bf R}^4$ is mapped to ${\bf R}\times S^3$. We compactify the ${\bf R}$ by making $\tau$ a  periodic variable. }
\begin{align}
(\vec{x}, x^4 ) = e^{-\tau} ( \sin \chi \vec{ \Omega}, \cos \chi), \quad  \left( \vec{\Omega} := ( \sin \theta \cos \varphi, \sin \theta \sin \varphi, \cos \theta) \right) \ . \nn
\end{align}
Then the metric of $S^3$ in $S^1 \times S^3$ is given as 
\begin{align} 
ds^2_{S^3} =  d \chi^2 + \sin^2 \chi ( d \theta^2 + \sin^2 \theta d \varphi^2 ), \label{3-sphere}
 \end{align}
where $ \chi\in [0, \pi]$, $ \theta \in [0, \pi]$ and $ \varphi \in [0, 2 \pi ]$. 

 In \cite{Dimofte:2011py}, the following superconformal index for 4d ${\cal N}=4$ SYM is shown to be useful to study domain wall or line operator indices
\begin{align}
I (x) &= {\rm Tr} [ (-1)^F x^{ \epsilon  + j_L + j_R}], 
\label{index}
\end{align}
which is the same index employed in \cite{Gang:2012yr} to study 4d line operator indices. $ \epsilon, j_L, j_R$ are Cartan generators of $U(1)^3 \subset U(1) \times SU(2)_L \times SU(2)_R \simeq SO(2) \times SO(4)$, which is the rotation symmetry of $S^1 \times S^3$. One can choose $j_L$, $j_R$ so that  $j_L+j_R$ rotates the phase of $x^1+ i x^2$ on ${\bf R}^4$, which in turn rotates $ \varphi$
 of the 3-sphere coordinates in \eqref{3-sphere}.

To consider the index with a domain wall, we need to consider a 3d index on the domain wall, since there are new degrees freedom localized at the 3d domain wall. One can use the following 3d ${\cal N}=2$ superconformal index on $S^1 \times S^2$, 
\begin{align}
I_{3d} (x) = {\rm Tr} [ (-1)^F x^{ \epsilon + j} ] ,  \label{3d-ind}
\end{align}
where $ j$ is the Cartan of $SO(3)$ rotation of $S^2$. Under the embedding of the domain wall ($S^1\times S^2 \subset S^1 \times S^3$), the $j$ corresponds to the $j= j_L + j_R$ in eq~\ref{index}. Thus  the 3d/4d index in eq~\eqref{3d-ind}/eq~\eqref{index} can be used for the 3d/4d coupled system.  This 3d index has been studied, for instance, in \cite{Kim:2009wb, Imamura:2011su, Kapustin:2011jm}. 

To couple the 3d theories to 4d theories with insertions of   't Hooft line operators,  we will consider 3d generalized index introduced in \cite{Kapustin:2011jm}. That means,  we will turn on background monopole fluxes  which couple to the global symmetry $G \times {}^L G$ in the 3d theory. 

\section{Index with Duality Domain Wall}

\subsection{Index Formula}

The index \eqref{index} for 4d  ${\cal N}=4$ SYM  with gauge group $G$ can be written in the following form \cite{Dimofte:2011py}
\begin{align} 
I_{{\cal N}=4 \textrm{ SYM}, \ G} (x) 
&= \sum_{{\bf m}}\int [ d U]_{ \bf m}  
( \Pi_{{\bf m}, G} (x,U))^{ \dagger}  \Pi_{{ \bf m}, G} (x,U)\;.
\label{4d-ind:long}
\end{align}
Here $\Pi_{{\bf m}, G} (x,U)$(or $\Pi_G (x, U, {\bf m})$) denotes the ``half-index'' introduced in \cite{Dimofte:2011py}. The half-index can be understood as the index on $S^1 \times D_3$ ($D_3$ denotes the hemisphere) with  boundary ($S^1 \times S^2$) conditions  labelled by the magnetic fluxes on $S^2$, ${\bf m}$, and gauge holonomy along the $S^1$,  $U$.
$[dU]_{ \bf m}$ is a natural measure introduced in \cite{Dimofte:2011py} for $G=SU(2)$, which can be easily extended to an arbitrary gauge group $G$. For convenience,  we introduce the operation `$\odot$' defined as 
\begin{align}
(A\odot B) (\ldots,\ldots) := \sum_{ \mathbf{m}} \oint [dU]_{\mathbf{m}} A(\ldots, U,{\bf m}) B(U,{\bf m},\ldots)\;.
\end{align}
Then,  the index formula \eqref{4d-ind:long} can be simply written as 
\begin{align}
I_{{\cal N}=4 \textrm{ SYM}, \ G} (x)= \Pi_G^\dagger \odot \Pi_G \;. \label{4d-ind}
\end{align}
For an explicit expression, first introduce a basis $\{H^i\}$ of Cartan subalgebra of gauge group $G$ and let 
\begin{align}
{\bf m} = \sum_{i=1}^{\textrm{rank}(G)} m_i H^i\; , \; U= e^{i \lambda} = e^{i \sum_i \lambda_i H^i}\;.
\end{align}
We choose the normalization for $H^i$ so that $U=e^{i\lambda}$ is periodic  in $\lambda_i$s with  period $2\pi$.  The half-index is given by 
\begin{align}
 \Pi_{{\bf m}, G} (x, U) &= \delta_{{\bf m},{\bf 0}} PE[ - \frac{ x^2}{1-x^2} \chi_{adj} (U) + \frac{ x}{1-x^2} \chi_{adj} (U)]\;,
\label{half-index}
\end{align}
where $PE$ denotes the Plethystic exponential which converts a single particle index to a multi-particle index
\footnote{$PE$ is defined as 
\begin{align}
PE[f(x, U)] := \exp \left( \sum_{n=1}^{ \infty} \frac{1}{n} f(x^n , U^n ) \right). \nn
\end{align}
},
and $ \chi_{adj} $ is the character of the adjoint representation of $G$. Explicitly,

\begin{align}
\chi_{adj}(U) = \sum_\alpha e^{i \alpha(\lambda)}\;,
\end{align}
where $\sum_\alpha$ denotes the sum over all roots of gauge group $G$.
 The first term in the exponent of \eqref{half-index} is originated from fermions in the vector multiplet, the second term from scalars in the adjoint hyper-multiplet. The measure $[dU]_{\bf m}$ is given by  ( `sym' denotes the symmetry factor)
\begin{align}
[dU]_{\bf m} := \frac{1}{{\rm sym}({\bf m})}   \big{(}\prod_{i=1}^{\textrm{rank}(G)} \frac{d\lambda_i}{2\pi} \big{)}  x^{- \frac{1}2\sum_\alpha |\alpha ( \bf{m})|} \prod_{\alpha\neq 0} (1-x^{|\alpha(\bf{m})|} e^{i \alpha (\lambda)})\;. \label{monopole measure}
\end{align}

Before considering the superconformal index with the duality domain wall, let us first review the $S^4$ partition function with the domain wall. 
The expectation value of the domain wall on $S^4$ is proposed in \cite{Drukker:2010jp} as a matrix integral
\begin{align} 
\langle \hat{O}_{wall} \rangle_{S^4} = \int d \nu (a^{ \prime}) d \nu (a) \bar{Z}_{{\rm inst}} (a^{ \prime}) Z_{S^3} (a^{ \prime}, a) Z_{{\rm inst}} (a), 
\label{wall-exp}
 \end{align}
where $a$ (resp. $a^{ \prime} $) is the Coulomb vev  of ${\cal N}=4$ SYM at, say,  the South (resp. North) pole of $S^4$. $ d \nu (a)$ is the integration measure with the 1-loop determinant for a 3d ${\cal N}=2$ vector multiplet.  $Z_{\rm inst}(a)$ (resp. $Z_{\rm inst} (a^{ \prime})$) is the Nekrasov instanton partition function with holomorphic coupling $ \tau$ localized at the South (resp. North) pole of $S^4$. $Z_{S^3}(a, a^{ \prime})$ is the $S^3$ partition function for the theory on the domain wall where $a, a^{ \prime}$ are now FI and mass parameters of the 3d theory. In \cite{Drukker:2010jp}, the $S^3$ partition function on the duality domain wall is conjectured to be equivalent to the duality kernel in the 2d Louville theory in the AGT context. The conjecture is explicitly checked in \cite{Hosomichi:2010vh}. 

On the other hand, without a domain wall, the partition function on $S^4$ is given by Pestun \cite{Pestun:2007rz}
\begin{align} 
Z_{{\cal N}=4 \textrm{ SYM}, \ S^4} = \int d \nu(a) \bar{Z}_{\rm{inst}} (a) Z_{\rm{inst}} (a). 
\label{partition}
\end{align}

An analogy between the instanton partition function and the half-index was established in \cite{Dimofte:2011py}, as can be seen in eq~\eqref{4d-ind} and eq~\eqref{partition}.
Using the same analogy, from eq~\eqref{wall-exp}, one can propose the S-duality domain wall index  as follows, 
\begin{align} 
&I_{wall} (x) \nonumber
\\
&=   \sum_{\bf m} \sum_{{\bf m}'}\int [ d U]_{\bf m} [dU^{ \prime}]_{{\bf m}'}  
\left( \Pi_{{\bf m}, G} (x, U) \right)^{ \dagger} I_{T[G]} (x, U, {\bf m}, U^{ \prime},{\bf m}^{\prime}) \Pi_{{\bf m}, {}^L G} (x,U^{ \prime})  \nn
\\
&= \Pi_G^\dagger \odot I_{T[G]} \odot \Pi_{{}^L G} \;.
\label{domain-ind}
\end{align}
where $I_{T[G]}(x, U, {\bf m},  U^{ \prime}, {\bf m}^{\prime})$ is the 3d index for $T[G]$ theory. $U$ and $U^{ \prime}$ (resp. ${\bf m}$ and ${\bf m}^{ \prime}$) are  chemical potentials (resp. magnetic fluxes) for global symmetries  $G$ and ${}^L G$. Since ${}^L G$ is realized as a quantum symmetry due to monopole operators, $U^{ \prime}$ will be introduced to the index of $T[G]$ as a chemical potential for the monopole charges. 

\subsection{Index on Duality Domain Wall  as a Duality Kernel}

Let us now argue that the S-duality wall index \eqref{domain-ind} should be same with the index for 4d ${\cal N}=4$ SYM {\it without} any domain wall.  Since an index is invariant under a continuous change of $\tau$, the S-duality wall index at any value of $ \tau$ should be same with the index evaluated at $ \tau = \frac{ i}{n_G} $ \footnote{$n_G =1$ for gauge group $G=SU(N)$, $SO(2N)$ and $n_G = \sqrt{2}$ for  $G=SO(2N+1)$, $USp(2N)$.}.  Consider an S-duality wall at $ \tau=\frac{i}{n_G}$.  This value is a fixed point of S-duality, i.e., $  \tau= -\frac{1}{n_G^2 \tau} = \frac{ i }{n_G}$. Thus taking  the inverse of S-duality on the right half-plane results in ${\cal N}=4$ SYM with $ \tau=\frac{i}{n_G} $ on the whole plane {\it without} any Janus wall. Therefore, the S-duality wall index given in eq~\eqref{domain-ind} should be same with the index for  4d ${\cal N}=4$ SYM in eq~\eqref{4d-ind}, 
\begin{align}
I_{wall} (x) = I_{{\cal N}=4 \textrm{ SYM}, \ G} (x) .
\label{trivial} \end{align}
We propose that the following sufficient condition for eq~\eqref{trivial} holds, 
\begin{align}
\Pi_{{\bf m}, \ G} (U) &= \sum_{{\bf m}^\prime}\int [dU^{ \prime}]_{{\bf m}^\prime} \  I_{T[G]} (U, {\bf m},  U^{ \prime} , {\bf m}^{\prime} ) \ \Pi_{{\bf m}^{\prime}, {}^LG} (U^{ \prime}) \   \nn
\\
 &=( I_{T[G]}\odot \Pi_{{}^LG} )_{\bf{m}}(U)\;.
\label{condition}
\end{align}
which will be explicitly checked for $G=SU(2)$ in section~\ref{sec: Check} and for $G=SU(3)$ in appendix \ref{app:2}. A similar relation is found for $S^4$ and $S^3$ partition functions  \cite{Hosomichi:2010vh}.

The relation \eqref{condition} can be further generalized by introducing BPS line operators to the system. As discussed in \cite{Dimofte:2011py}, if half-BPS Wilson line and 't Hooft line operators preserve  the rotation symmetry generated by $j= j_L+j_R$, they preserve at least 2 real super charges which are common to that preserved by a half-BPS domain wall at $ \chi = \frac{ \pi}{2}$. The line operators on $S^1 \times p \subset S^1 \times S^3$ preserve $j$, if the point $p$ is on $S^2 \subset S^3$ defined by $ \theta=0$ or $ \theta = \pi$ in \eqref{3-sphere}. The preserved supersymmetries are compatible with the definition of the superconformal index in \eqref{index} and \eqref{3d-ind}. Thus one can consider the index in the presence of both the duality domain wall and the line operators. With the insertion of a line operator $L$, the half-index $\Pi$ is modified as 
\begin{align}
\Pi \qquad   \stackrel{\textrm{+ line operator $L$}}{\rightarrow}  \qquad \hat{O}_{L}\cdot \Pi\;.
\end{align}
where $ \hat{O}_L$ is a difference operator acting on the half-index $\Pi$. 
\begin{figure}[h]
\begin{center}
   \includegraphics[scale=0.45]{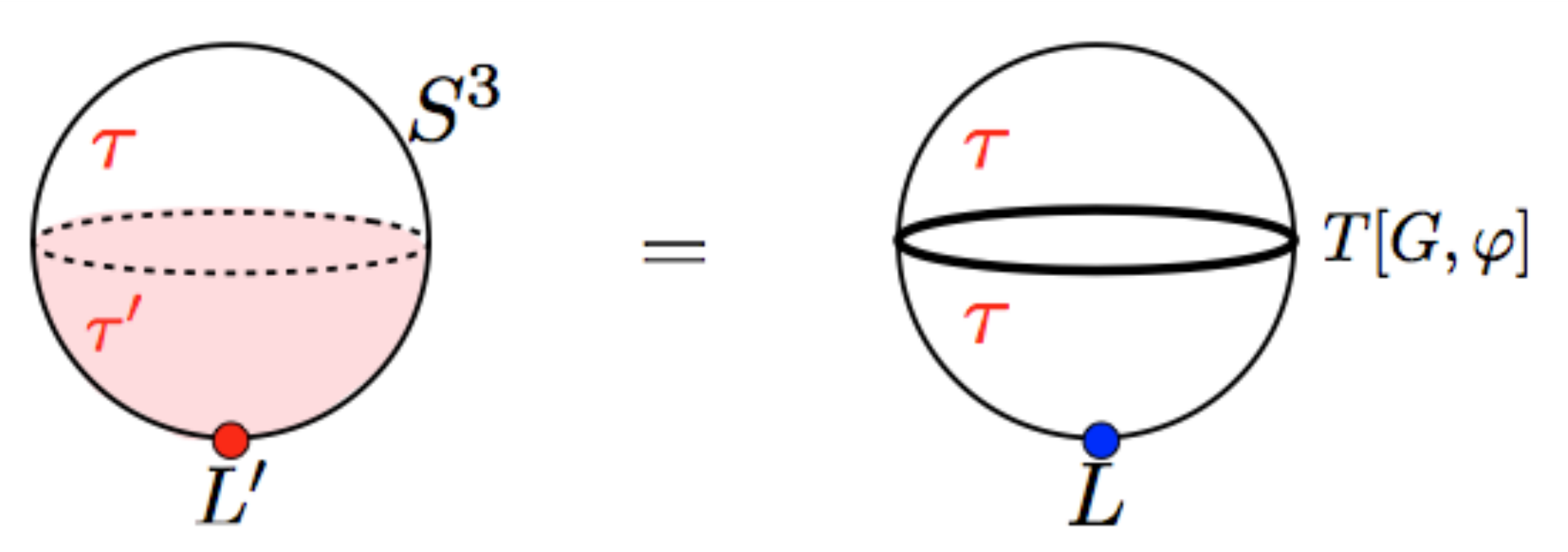} 
   \end{center}
   \vspace{-0.8cm}
   \caption{Consider a Janus domain wall at the equator of 3-sphere in $S^3 \times S^1$  with a line operator $L^{ \prime}$ inserted at the south pole of the 3-sphere (left).  $ \tau^{ \prime}$ and $L^{ \prime}$ are related with $ \tau$ and $L$ by a duality element $ \varphi$. Taking $ \varphi^{-1}$-daulity on the southern hemisphere results a duality domain wall at the equator and the line operator $L$ at the south pole (right). }
\end{figure}

Generalizing eq~\eqref{condition} including line operators, we propose the following  
\begin{align} 
  & I_{T[G,\varphi]} \odot (\hat{O}_L \cdot \Pi) = \hat{O}_{\varphi(L)}\cdot \Pi
\label{s-dual-line}
\end{align}
where $\varphi(L)$ denotes a line operator related with $L$ by $ \varphi \in PSL(2, Z)$.  We will show explicit examples for eq~\eqref{s-dual-line} for $G=SU(2)$ in section~\ref{sec: Check}.

\section{Example :  $\mathcal{N}=4$ $SU(2)$ SYM}

\subsection{Index for $T[SU(2),\varphi]$} \label{sec: T[SU(2)]}
In this section, we will write down the explicit formula for generalized index for  $T[SU(2),\varphi]$ with general $\varphi$. One can use the prescription in \cite{Kim:2009wb,Imamura:2011su,Kapustin:2011jm} to write down the 3d generalized index for $T[SU(2)]$ theory as follows 
\begin{align}
&I_{T[SU(2)]} (  u, m  ;  u^{ \prime}, m^{ \prime} ) =  \sum_{  s}^{}
\oint    \frac{ d\zeta}{ 2 \pi i \zeta}    \left( u^{ \prime} \right)^{2 s}  \zeta^{2  m^{ \prime} } x^{\frac{1}{2} |s+m|+\half |s-m| } \nn \\
&\qquad \times PE[   \frac{ x^{1/2}}{1+x} \left( x^{ | s+m |}  ( u \zeta+  u^{-1} \zeta^{-1})+  x^{|s-m|} (u \zeta^{-1} +u^{-1} \zeta ) \right) ]\;.
\label{tsu2} 
\end{align}
Recall that $T[SU(2)]$ theory is given by 3d $\mathcal{N}=4$ SQED with two electron hypermultiplets. Here $U={\rm diag}(u, u^{-1}), U^{ \prime}={\rm diag}(u^{ \prime}, {u^{ \prime}}^{-1})$ are chemical potentials for (or equivalently background gauge holonomies coupled to) the global symmetry $SU(2)$  and  ${}^L SU(2)$ respectively. Similarly $\mathbf{m}={\rm diag} (m, -m)$, $\mathbf{m}^\prime ={\rm diag}(m^\prime, -m^\prime)$ are background monopole fluxes coupled to the global symmetries.  ($\zeta,s$) are the holonomy and magnetic fluxes of the (dynamical) $U(1)$ gauge group respectively. $x^{ \half |s+m|+ \half |s-m|}$ arises from the Casimir energy of the monopole sector. The term $(u^{\prime})^{ 2s}  \zeta^{2m^{ \prime}}  $ comes from  the coupling of the background gauge field $A_{BG}$ with the topological current $J=*dA_{U(1)}$ \cite{Kapustin:2011jm}, 
\begin{align}
\int   A_{BG}\wedge dA_{U(1)}\;.
\end{align}
The origin of terms in the Plethystic exponential can be found in  \cite{Imamura:2011su}. The sum  $ \sum_s$ is over integers (resp. half-integers) if $m$ is an integer (resp. half-integer). It is to satisfy the Dirac quantization conditions, $ s \pm m \in \mathbb{Z}$. 
Note that  $ \zeta, u, u^{ \prime} $ are complex numbers with the unit length and their complex conjugations are given by $ \zeta^{*}= \zeta^{-1}$, etc. Expanding in $x$, the first two terms are 
\begin{align} 
I_{T[SU(2)]} (u, m=0 ; u^{ \prime},m^{\prime}=0 ) &=
1+( \chi_{adj} ( u) + \chi_{adj} (u^{ \prime})) x  + O(x^2) 
\nn 
\end{align}
where $ \chi_{adj}(u)= u^{-2} + 1+ u^2 $ is the adjoint character for $SU(2)$. 
Thus up to this order, it is obvious that the index is invariant under the exchange of $(u,m)$ and $(u^{ \prime},m^{\prime})$ 
\begin{align} 
I_{T[SU(2)]} (u, m ; u^{ \prime},m^{\prime}) =  I_{T[SU(2)]} (u^{ \prime},m^{\prime} ;  u , m)\;. \label{self-dual-tsu2}
 \end{align}
Indeed, eq~\eqref{self-dual-tsu2} is checked to several orders in $x$ using Mathematica with different values of $m$ and $m^{ \prime}$. 
 For instance, 
\begin{align} 
& I_{T[SU(2)]}  (u, 0 ;  u^{  \prime}, \half) = I_{T[SU(2)]}  (u^{  \prime}, \half ;  u, 0)  \nn \\
& \quad = (u^{-1}+u ) x^{1/2} + (u^{-3}+u^{-1} + u+ u^3 ) x^{3/2} + \ldots \;,  \textrm{ and}
\nn \\
&  I_{T[SU(2)]}  (u, 1 ;   u^{  \prime}, 3) = I_{T[SU(2)]}  (u^{  \prime}, 3 ;  u, 1) \nn \\
& \quad =  (u^{-6} {u^{ \prime}}^{-2} + u^6 {u^{ \prime}}^2 ) x^4 + (u^4 {u^{ \prime}}^2 + u^{-4} {u^{ \prime}}^{-2} ) x^6 + \ldots 
\end{align}
$T[SU(2)]$ theory is known to be self-dual under the 3d mirror symmetry \cite{Intriligator:1996ex}.
 Eq~\eqref{self-dual-tsu2}  shows the self-mirror property of the superconformal index of $T[SU(2)]$, especially the interchangeability of Coulomb and Higgs branch. A similar self-mirror property of $S^3$ partition function of $T[SU(2)]$ is noted in \cite{Hosomichi:2010vh}. 

  Let us now consider an arbitrary element $\varphi$ in $PSL(2, \mathbb{Z})$.  $\varphi$  can be written in the following form 
\begin{align}
\varphi =\ldots \cdot (ST^{k_2}) \cdot (ST^{k_1}) \;. \label{Decomposition of PSL(2,Z)}
\end{align}
For $\varphi = ST^k$, the index can be given as
\begin{align}
I_{T[SU(2);\varphi = ST^k]} (u,m;u^\prime,m^\prime ) =(u^{\prime})^{2km^\prime} I_{T[SU(2)]}(u,m;u^\prime, m^\prime) \;.
\end{align} 
since the $T^k$ action corresponds to adding a  Chern-Simons term with level $k$ for background gauge field which couples to the quantum $SU(2)$ symmetry (see fig.~\ref{fig:quiver}).
For $\varphi = \varphi_2 \cdot \varphi_1$, the corresponding index is given as
\begin{align}
&I_{T[SU(2),\varphi = \varphi_2 \cdot \varphi_1]}(u,m ; u^\prime, m^\prime) \nn
\\
&= \sum_{n=0,\frac{1}2,1,\ldots}^\infty \oint [dv]_n I_{T[SU(2),\varphi_2]}(u,m;v,n)I_{T[SU(2),\varphi_1]} (v,n;u^\prime,m^\prime)\;, \nn
\\
&= \big{(} I_{T[SU(2), \varphi_2]}\odot I_{T[SU(2),\varphi_1]} \big{)}(u,m ; u^\prime, m^\prime) \;.
\label{index for STk}
\end{align}
As we  see in  fig.~\ref{fig:quiver2},   the 3d theory $T[SU(2), \varphi_2 \cdot \varphi_1]$ is obtained by gluing ${}^L SU(2)$   in $T[SU(2), \varphi_2]$  and $SU(2)$   in $T[SU(2), \varphi_1]$. In the gluing we  gauge the diagonal part of the two $SU(2)$s and this gauging results in the integration (summation) over holonomy $v$ (flux $m$) in the above index formula.  Here, the measure $[dv]_n$ is given as \eqref{monopole measure}
\begin{align}
[dv]_n =(1- \frac{1}2 \delta_{n,0})  \frac{d v}{2 \pi i v } x^{- 2|n|} (1-v^2  x^{2 |n|})(1-v^{-2}x^{2 |n|}) \;.
\label{Index multiplication}
\end{align}
Combining \eqref{tsu2}, \eqref{index for STk} and \eqref{Index multiplication},  one can write down the index formula  for $T[SU(2),\varphi]$ with arbitrary $PSL(2,\mathbb{Z})$ element $\varphi$. 

Note that for given $\varphi \in PSL(2,\mathbb{Z})$, the way of decomposing $\varphi$ as products of  $ST^k$s  \eqref{Decomposition of PSL(2,Z)} is not unique. For example,  $\varphi$ can be expressed as
\begin{align}
\varphi =S^2 \cdot \varphi \;.
\end{align}
Each decomposition of $\varphi$ will give a different formula for the index $I_{T[SU(2),\varphi]}$. But interestingly they seem to give the  same index. For example, under the  decomposition $ S^2 \cdot\varphi $,  
the index can be written as
\begin{align}
&I_{T[SU(2),   S^2  \cdot\varphi]} = I_{T[SU(2), S^2]} \odot I_{T[SU(2), \varphi]}\;.
\end{align}
As argued in Appendix \ref{sec:I[T[SU[2],S2]]}, $I_{T[SU(2),S^2]}$ acts as the identity operator on 3d index, i.e., 
\begin{align}
I_{T[SU(2),S^2]} \odot I_{T[SU(2), \varphi]} = I_{T[SU(2), \varphi]} \;.
\end{align}
Thus for this simple case, we confirm that the index $T[SU(2), \varphi]$ does not depend on the decomposition of $\varphi$ ($\varphi$ or $S^2 \cdot \varphi$). 
It would be interesting to prove it generally that the index  $I_{T[SU(2),\varphi]}$ does not depend on the decomposition of $\varphi$.   
 
 The property \eqref{self-dual-tsu2} which holds only for $\varphi =S$  case can be  generalized to general mapping class element $\varphi \in PSL(2,\mathbb{Z})$ as follows
\begin{align}
I_{T[SU(2),\varphi^{-1}]} (u,m;u^\prime,m^\prime) = I_{T[SU(2),\varphi]} (u^\prime,-m^\prime;u,-m)\;. \label{generalization of self-dual-tsu2}
\end{align}
For $\varphi=S$ case, $\varphi^{-1}=\varphi$ as an element of $PSL(2,\mathbb{Z})$ and $I_{T[SU(2)]} (u,m;u^\prime,m^\prime)=I_{T[SU(2)]} (u,-m;u^\prime,-m^\prime)$.  
Eq.~\eqref{generalization of self-dual-tsu2} can be checked by explicit index calculations. For example, 
\begin{align}
&I_{T[SU(2),\varphi^{-1}=(ST)^2]} (u,\frac{1}2;u^\prime,\frac{3}2) = I_{T[SU(2),\varphi=ST]} (u^\prime,-\frac{3}2;u,-\frac{1}2)\;, \nn
\\
& = (u^2 u^\prime+\frac{1}{u^4 u^\prime})x^2 + (u^\prime + \frac{1}{u^2 u^\prime}) x^3\ldots
\end{align}
It is worthy to prove the property in eq.~\eqref{generalization of self-dual-tsu2} in full generality.

 \subsection{Half-index and Line Operators} \label{sec:line}
In this section, we will give the explicit expression for the half-index $\Pi_m$ in eq.\eqref{half-index} and the action of line operators $\hat{O}_L$ on the half-index
for $G=SU(2)$. Firstly, 
\begin{align}
 \Pi_m (x, u)
&= \delta_{m,0} PE[  \frac{ x}{1+x} \chi_{adj}(u)] .\label{half-ind-su2}
\end{align}
Line operators  in the theory are labelled by two integers $(p,q)$. Using the Weyl transformation of the gauge group $SU(2)$ which relates $(p,q) \sim (-p, -q)$, one can assume that $ p \geq 0$.  Under $S$ and $T$ transformation of $PSL(2,\mathbb{Z})$,  line operators $(p,q)$  of magnetic charge $p$ and electric charge $q$   transform as follows 
\begin{align}
S: (p,q) \mapsto (-q, p), \qquad T : (p, q ) \mapsto ( p , p+q). 
\label{st}
\end{align}
Basic line operators are ${\cal W}$ (the fundamental Wilson line), ${\cal T}$ ('t Hooft line operator with the minimal magnetic charge)  and ${\cal D}$ (a dyonic line operator with minimal electric/magnetic charge). They correspond to $(p,q)=(0,1),(1,0)$ and $(1,1)$ respectively. These three are not independent but constrained with the following relations \cite{Dimofte:2011jd} , 
\begin{align}
&x^{1/2} {\cal W T} - x^{-1/2}{\cal T W } = (x-x^{-1}){\cal D}   \;, \nn
\\
&\textrm{and its cyclc permuations $ {\cal W}  \to {\cal D} \to {\cal T} \to {\cal W} $}\;.
\label{line relations}
\end{align}
Under the $S$ and $T$, these line operators transform as
\begin{align}
&S \;:\; {\cal W}\;\rightarrow \;{\cal T}\;, \quad {\cal T}\;\rightarrow {\cal W}\;, \quad {\cal D} \rightarrow\; -x {\cal D}+x^{\half} {\cal WT}\;, \nn
\\
&T \;:\; {\cal W}\;\rightarrow \;{\cal W}\;, \quad {\cal T}\;\rightarrow {\cal D}\;, \quad {\cal D} \rightarrow\; - x {\cal T} + x^{ \half} {\cal DW}  \; . \label{s-map}
\end{align}
Note that these transformation rules are  compatible with the relations in eq.~\eqref{line relations}. 

The half-index with a line operator $(p,q)$ can be obtained by acting a difference operator $\hat{O}_{p,q}$ on the half index $\Pi$. 
Explicit forms of difference operators $\hat{O}_{1,s}$ and $ \hat{O}_{0,1}$ are given as \cite{Dimofte:2011py}
\begin{align} 
\hat{O}_{1,s} &= \frac{\hat{x} x^{ - \half} - \hat{x}^{-1} x^{ \half} }{ \hat{x} - \hat{x}^{-1} } x^{ - \frac{s}{2} } \hat{x}^s \hat{p}^{ - \half} + \frac{ \hat{x} x^{ \half} - \hat{x}^{-1} x^{ - \half} }{ \hat{x} - \hat{x}^{-1} } x^{ - \frac{s}{2} } \hat{x}^{-s } \hat{p}^{ \half} \ , 
\nn 
\\
\hat{O}_{0,1} &= \hat{x}+ \hat{x}^{-1} \ , 
\label{explicit} 
\end{align}
where the basic difference operators $\hat{x}$ and $\hat{p}$ are
\begin{align} 
\hat{x} : & = x^m u , \qquad 
\hat{p} : = e^{ \partial_m }  x^{   \frac{ \partial}{ \partial_{ \ln u} }}. 
\nn \end{align}
By their definitions,  $\hat{O}_{{\cal W}} :=\hat{O}_{0,1}, \hat{O}_{{\cal T}}:=\hat{O}_{1,0}$ and $\hat{O}_{\cal D} :=\hat{O}_{1,1}$.
One can check that
\begin{align}
&x^{1/2} \hat{O}_{\cal W} \hat{O}_{\cal T} - x^{-1/2} \hat{O}_{\cal T} \hat{O}_{\cal W}  - (x-x^{-1})\hat{O}_{\cal D} =0\; \nn
\\
&\textrm{and its cyclc permuations $ {\cal W}  \to {\cal D} \to {\cal T} \to {\cal W} $}\;.
\end{align}
These are  consistent with eq.~\eqref{line relations}.

Let us consider how to obtain an explicit form of $\hat{O}_{p,q}$. Under a $PSL(2,\mathbb{Z})$ transformation, the greatest common divisor of $|p|$ and $|q|$ is invariant. Since $\gcd(n,0)=n$ for $n \in {\mathbb Z}_{+} $, the $PSL(2, \mathbb{Z})$ orbits of $\hat{O}_{n,0}$ are different for each other value of $n$. For any given charge $(p,q)$ of a line operator, one can do the following procedure repeatedly 
\begin{enumerate}
\item Apply $T$ (or $T^{-1}$) transformations until the absolute value of electric charge becomes less than or equal to the magnetic charge. 
\item Apply $S$ transformation. 
\end{enumerate}
to relate $\hat{O}_{p,q}$ with $\hat{O}_{n,0}$ by a $PSL(2, {\mathbb Z})$ transformation where $n=\gcd(|p|,|q|)$. Thus $PSL(2, \mathbb{Z})$ orbits of line operators $\hat{O}_{p,q}$ can be classified by a non-negative integer, $\gcd(|p|, |q|)$. In \cite{Gomis:2011pf}, it has been noted that the line operator $(\hat{O}_{1,0})^n$ corresponds to $\hat{O}_{n,0}$. Given $S$ and $T$ transformation of basic line operators in eq~\eqref{s-map}, one can generate an explicit form of $\hat{O}_{p,q}$ for any $(p,q)$ using eq~\eqref{explicit}. 

The $PSL(2,\mathbb{Z})$ transformation rules \eqref{s-map}  and algebraic relations \eqref{line relations} between line operators   are obtained in \cite{Dimofte:2011jd} by studying  Teichm$\ddot{\rm u}$ller space of the one-punctured torus. They  use the relation between the shear coordinates and loop coordinates of the quantum Teichm$\ddot{\rm u}$ller space, the relation between loop coordinates  and loop operators in ${\cal N}=4$ theory, and the duality transformation of the shear coordinates. The transformation given in eq.(3,45) of \cite{Dimofte:2011jd} is identical to our eq~\eqref{s-map} up to some relative signs, by relating $(\hat{x}, \hat{y}, \hat{z})_{theirs} = ({\cal T, W, D})_{ours}$. The quantum parameter $ \hbar$ is related with their $q$ as $(q)_{theirs}=e^{ \hbar}$ and with our $x$ as $ (x)_{ours} = (q)_{theirs}^{ \half}$  \cite{Dimofte:2011py, Gang:2012yr}. It's noteworthy to mention that   geometric structures of a 2d Riemann surface  $\Sigma$ are  encoded in the superconformal indices (not only in the $S^4$ partition function \cite{Drukker:2009id})
of a  4d theory obtained by compactifying the $6d$ $(2,0)$ theory on $\Sigma$.

\subsection{Checks for eq.~\eqref{s-dual-line}} \label{sec: Check}
In this section, we will check eq.~\eqref{s-dual-line} with various examples in the series expansion in $x$. 

For $L={\cal T}^l {\cal W}^n$ and $\varphi= S$,  the eq.~\eqref{s-dual-line} becomes
\begin{align} 
& \sum_{m^{ \prime}} \oint [ d u^{ \prime} ]_{m^{ \prime}}  I_{T[SU(2)]} ( u , m ; u^{ \prime}, m^{ \prime} ) ( \hat{O}_{1,0} ^{l} \hat{O}_{0,1}^{n} ) \cdot \Pi_{m^{ \prime}} (u^{ \prime})   = ( \hat{O}_{0,1}^l \hat{O}_{1,0}^n ) \cdot \Pi_{m} (u) \;. \label{w-to-t}
\end{align}
Using the explicit expressions  in the previous sections (4.1, 4.2),  one can check these. For example, consider $l=n=0$ and $m=0$. Using the fact that $\Pi_{m^\prime}(u^\prime)$ is zero unless $m^\prime=0$,  the above equation is simplified as  
\begin{align}
\Pi_{m=0} (x, u) &= \oint [du^{ \prime}]_{m^\prime=0} \  I_{T[SU(2)]} (x;  u, m=0, u^{ \prime} ,m^\prime=0) \ \Pi_{m^\prime=0} (x, u^{ \prime}) \;.\label{cond-tsu2} 
\end{align}
 One can check that both sides give the same series expansion in $x$, that is  
 \begin{align}
& 1+(u^{-2} +1+u^2 ) x + (u^{-4}+1+u^4) x^2 + (u^{-6} + u^{-2} + u^2 + u^6 ) x^3 \nn \\
& \quad + (u^{-8} + u^{-4} +2+ u^4 + u^8) x^4+ O(x^5) \;. \nn 
\end{align}
For $L={\cal T}^n$ and $\varphi= ST$,  $ \varphi (L) = (- x {\cal D} + x^{ \half} {\cal W T})^n $, thus 
\begin{align}
&\sum_{m^{ \prime}} \oint [ d u^{ \prime} ]_{m^{ \prime}}  I_{T[SU(2);\varphi=ST]} ( u , m ; u^{ \prime}, m^{ \prime} )  \hat{O}^n_{1,0}  \cdot \Pi_{m^{ \prime}} (u^{ \prime}) \nn
\\
&= ( - x\hat{O}_{1,1}+ x^{ \half} \hat{O}_{0,1} \hat{O}_{1,0} )^n \cdot \Pi_{m} (u) \; , 
\end{align}
In the similar way, perturbatively in $x$,  the above equation can be checked for an arbitrary $n$.

\subsection{Mass-deformation to ${\cal N}=2^*$ theory}
In this section, we will consider the mass deformation of 4d ${\cal N}=4$ theory, namely $\mathcal{N}=2^*$ theory. For the sake of simplicity, we will turn off the background monopole fluxes in this section, $m=m^\prime=0$. Turning on the background monopole flux is rather straightforward.

One can deform the 4d ${\cal N}=4$ theory to ${\cal N}=2^*$ theory by turning on a mass parameter for the hyper-multiplet. 
 The half-index for 4d theory differs from \eqref{half-ind-su2} as
 \begin{align} 
 \Pi (u, \eta) &= PE[ \left( - \frac{ x^2}{1-x^2}+ \frac{ x}{1-x^2} \eta \right)  (u^2 +1+u^{-2})] 
 \nn 
 \end{align}
 where $\eta$ is a chemical potential for $U(1)$ rotating the phase of the adjoint hyper-multiplet  \cite{Dimofte:2011py}. In the path-integral approach of the index, turning on a chemical potential corresponds to a twisting of the covariant derivative of $S^1$ direction. In this approach, it is clear that turning on $\eta$ corresponds to the mass deformation of ${\cal N}=4$ theory with the adjoint hyper-multiplet mass $  \frac{ \ln ( \eta ) }{ 2 \pi i } $ \cite{Gang:2012yr}. 

The mass deformation of 4d ${\cal N}=4$ SYM induces a deformation of $T[SU(2)]$ theory where the deformed Lagrangian is given explicitly in \cite{Hosomichi:2010vh}. In the superconformal index, 
it corresponds to turning on a chemical potential of $U(1)$ under which 4 complex scalars in the fundamental hyper-multiplets have a charge $1$, 
and the scalar chiral field in the vector-multiplet has a charge $-2$. This symmetry is called $U(1)_{\rm puncture}$ in \cite{Dimofte:2011py}. 
Using generalized index in \cite{Kapustin:2011jm}, we turn on the chemical potential $ \eta$ of  $U(1)_{\rm puncture}$ as follows
\begin{align}
 I_{T[SU(2)]} ( u, u^{ \prime},  \eta )& =    \sum_{  s =-\infty}^{\infty} \oint    \frac{ d\zeta}{ 2 \pi i \zeta}    \left( u^{ \prime} \right)^{2 s} \eta^{ - |s|} x^{|s| } 
\nn \\
&  \times    PE[   \frac{ x^{\frac{1}{2}}\eta^{\frac{1}{2}}-x^{\frac{3}{2}} \eta^{-\frac{1}{2}}}{1-x^2} x^{ | s|} ( \zeta+ \zeta^{-1}) (u+u^{-1})+ \frac{x \eta^{-1}- x \eta }{1-x^2}]. 
\label{tsu2-mass}
\end{align}
Note that the fermions are charged oppositely under $U(1)_{\rm puncture}$  from the scalars in the same chiral multiplet. $\eta^{-|s|}$ is originated from the zero-point contributions to the charge   of $U(1)_{\rm{puncture}}$
\cite{Imamura:2011su}. 
The first term in the plethystic exponential comes from the the 2 fundamental hyper-multiplets, 
while
the second term comes from the vector multiplet. Here we use the fact that the chiral field in an ${\cal N}=4$ vector-multiplet has a non-canonical R-charge $1$ \cite{Kapustin:2010xq}. 

The $U(1)_{\rm puncture}$ is an anti-diagonal sum of $U(1)$ subgroups of $SU(2) \times SU(2)$ R-symmetry of the undeformed ${\cal N}=4$ theory \cite{Tong:2000ky}
where two $SU(2)$s are exchanged under the 3d mirror symmetry.  Thus, under the mirror symmetry, $ \eta$  is mapped to the inverse of itself.  
The self-mirror property eq~\eqref{self-dual-tsu2} can be rewritten as 
\begin{align} 
I_{T[SU(2)]}( u, u^{ \prime}, \eta) = I_{T[SU(2)]} ( u^{ \prime}, u, \eta^{-1} ), 
\label{mir}
\end{align}
and it holds as high an order in $x$ as we checked. Indeed  eq~\eqref{mir} is proved analytically in \cite{Krattenthaler:2011da}. 

The argument that the index for the bulk theory should be same with the domain wall index still holds true after the mass deformation. 
The following relation, corresponding to the mass deformation of eq~\eqref{cond-tsu2}, is checked to several orders in $x$ using Mathematica, 
\begin{align}
\Pi (u, \eta)  &= \oint [du^{ \prime}] \  I_{T[SU(2)]} ( u, u^{ \prime}, \eta ) \ \Pi (u^{ \prime}, \eta )      \nn
\end{align}
where $[du^{ \prime}]:= [du^{ \prime}]_{m^{ \prime}=0}$. 

\subsection{Another description for $T[SU(2)]$ theory? }
In \cite{Teschner:2012em}, the authors have found a 3d theory whose squashed three sphere partition function is same with that of  $T[SU(2)]$ theory. 
In this section, we will compare the superconformal indices for both theories. 
The  theory found in \cite{Teschner:2012em} is 3d ${\cal N}=2$ $SU(2)$ Chern-Simons theory of level $k=1$ with  four fundamental chiral multiplets and three neutral chiral multiplets. The 3d theory has $SU(2)\times ^{L}SU(2) \times U(1)_{\textrm{puncture}}$  global symmetries which are same with that of $T[SU(2)]$.  The charges of chiral fields under the global symmetry can be read from expression of the partition function given in  \cite{Teschner:2012em}. 
 In table~\ref{tab:dual-tsu2},  we summarize charge assignment of chiral fields under the three Cartans, denoted as $U(1)_{\rm bot} \times U(1)_{\rm top} \times U(1)_{\rm puncture}$, of the global symmetries.
\begin{table}[h]
   \centering
   \begin{tabular}{@{} l | ccc   @{}} %
      \toprule
$SU(2)$     & $U(1)_{\rm bot}$ & $U(1)_{\rm top}$ & $U(1)_{\rm puncture}$   \\
      \midrule
      ${\bf 2}$ & 1 & 1 &  $  \frac{1}{2}$  \\
      \cmidrule{2-4} 
      & 1 & -1 & $  \frac{1}{2}$  \\
        \cmidrule{2-4} 
      & -1 & 1 & $ \frac{1}{2}$ \\
        \cmidrule{2-4} 
      & -1 & -1 & $ \frac{1}{2}$ \\
      \midrule
       \cmidrule{2-4} 
     1  & 0 &-2 & -1  \\
     \cmidrule{2-4}
     & 0 & 2 & -1 \\
          \cmidrule{2-4}
     & 0 & 0 & -1 \\
      \bottomrule
   \end{tabular}
   \caption{Charges of chiral fields in a theory found to be dual to $T[SU(2)]$ in \cite{Teschner:2012em}. $SU(2)$ in the first column denotes the gauge group. 
   }
   \label{tab:dual-tsu2}
\end{table}
The single particle index of the Chern-Simons theory can be written as 
\begin{align}
I_{CS}^{ sp} (x, u, u^{ \prime}, \eta ; \zeta)
= & 
(u+u^{-1})(u^{ \prime}+{u^{ \prime}}^{-1}) \frac{x^{ \frac{1}{2}} \eta^{ \frac{1}{2}} - x^{ \frac{3}{2}} \eta^{- \frac{1}{2}}}{1-x^2} x^{|s|} (\zeta + \zeta^{-1})
\nn  \\
& + \frac{x}{1-x^2} ( \eta^{-1} - \eta ) ( {u^{ \prime}}^2 + 2 + {u^{ \prime}}^{-2})
\end{align}
where the first line is originated from the four fundamental chiral multiplets and the second line from the neutral chiral multiplets.
The R-charge  for fundamental (resp. neutral) chiral fields is assigned to be $\frac{1}{2}$ (resp. $1$).  $(\zeta, s)$ are 
fugacity and magnetic flux for $SU(2)$ gauge group. 
The index of the Chern-Simons theory can be obtained by taking Plethystic exponential of the single particle index, 
\begin{equation}
I_{CS} ( u, u^{ \prime}, \eta)= \sum_{s=0}^{ \infty} \oint [ d \zeta]_s \eta^{ - 2 |s|} \zeta^{ 2 |s|} 
PE [ I^{sp}_{CS} ( x, u, u^{ \prime}, \eta;  \zeta) ] 
\label{ind:cs}
\end{equation}
where $[d \zeta]_s$ is the $SU(2)$ Haar measure in eq~\eqref{Index multiplication}.
 In eq~\eqref{ind:cs}, $ \zeta^{ 2 |s|} $ is originated from the classical contribution of Chern-Simons term
 and $ \eta^{ - 2 |s|} $ from the zero-point contribution to the flavor charge. 
Up to several orders in $x$, we check that 
\begin{equation}
I_{CS} (u, u^{ \prime}, \eta) = I_{T[SU(2)]} ( u, u^{ \prime}, \eta)
\end{equation}
for $I_{T[SU(2)]} ( u, u^{ \prime}, \eta)$ given in eq~\eqref{tsu2-mass}.

\section{Generalization to $A_1$ Gaiotto theories }
 One may consider a more general set-up. Consider Gaiotto theories \cite{Gaiotto:2009we}  obtained by compactifying the 6d $A_{N-1}$  (2,0) theory on a 2d surface $\Sigma_{g,h}$, a genus $g$ Riemann surface with $h$ holes. In this section, we will only consider the $N=2$ (two $M5$s) case and denote the corresponding 4d $\mathcal{N}=2$ superconformal theories by $ T_{g,h}$. Basic dictionaries between structures on the 2d surface and 4d field theory are as followings,
 \begin{itemize}
\item Space of complex structures on $\Sigma$ = Parameter space of the 4d theory.
\item Mapping class group of $\Sigma$ = (Generalized) S-duality group.
 \end{itemize}
 Let us denote the mapping class group of the $\Sigma_{g,h}$ by $\Gamma(\Sigma_{g,h})$.  For given Riemann surface $\Sigma_{g,h}$ and an element $\varphi \in \Gamma(\Sigma_{g,h})$, one can define 3d theory, denoted by $T[\Sigma_{g,h},\varphi]$, living on the duality domain wall between $T_{g,h}$ and $\varphi (T_{g,h})$. In this notation,
 \begin{align}
T[\Sigma_{1,1}, \varphi =S] = T[SU(2)]\; .	
\end{align}
Generalizing the formula in eq~\eqref{s-dual-line}, we expect that
\begin{align}
&\sum_{m^\prime}\oint [du^\prime]_{m^\prime} I_{T[\Sigma_{g,h},\varphi]} (u,m;u^\prime,m^\prime)\hat{O}_{L}\cdot \Pi_{m^{ \prime}} (u^\prime; T_{g,h}) \nn
\\
& = \hat{O}_{\varphi(L)}\cdot \Pi_{m}\big{(}u; \varphi(T_{g,h}) \big{)}\;. \label{general formula}
\end{align}
In general, we do not know the 3d theory $T[\Sigma_{g,h},\varphi]$, thus the index for the theory cannot be calculated from the prescription of  \cite{Kim:2009wb},\cite{Imamura:2011su},\cite{Kapustin:2011jm}. Thus we cannot check the above equation by directly calculating both sides. 
Instead, the above equation can be used as a tool to calculate the index for the mysterious 3d theory $T[\Sigma_{g,h}, \varphi]$. To determine the index $I_{T[\Sigma_{g,h}, \varphi]}$ from eq~\eqref{general formula}, we should know followings
\begin{enumerate}
\item $\Pi(T_{g,h})$, half-index  for the 4d theory $T_{g,h}$ .

\item $\hat{ \mathcal{L}}_{g,h}$, space of line operators in $T_{g,h}$ .

\item $\varphi(L)$, action of $\varphi$ on the line operators . 

\item $\hat{O}_L$, action of a line operator $L \in \hat{\mathcal{L}}$ on the half index \;.	
\end{enumerate}
Since we know the Lagrangian description for general $T_{g,h}$, given as $SU(2)$ quiver theories  \cite{Gaiotto:2009we},  we can easily calculate the half index. $\hat{ \mathcal{L}}_{g,h}$ and the action of $\varphi$ on line operators are studied in \cite{ Drukker:2009tz, Drukker:2009id}. Obtaining $\hat{O}_L$ can be rather
difficult but we succeeded for some examples in \cite{Gang:2012yr} which could be extended to general $SU(2)$ quiver theories.

\subsection{Example : ${\cal N}=2$ $SU(2)$ with $N_F=4$}

The 4d theory obtained from two M5 branes wrapping on $\Sigma_{0,4}$ is $\mathcal{N}=2$ $SU(2)$ gauge theory with 4 fundamental hypermultiplets, $N_F=4$. The half-index for the theory is given by 
\begin{align}
\Pi_m (T_{0,4}) (u)= \delta_{m,0} PE \big{[} \frac{4x}{1-x^2} (u+u^{-1})- \frac{x^2}{1-x^2}(u^2 +1+u^{-2})\big{]} \;.
\end{align}
For the fundamental Wilson line operator $L= {\cal W}$, the difference operator $\hat{O}_L$ is given by 
\begin{align}
\hat{O}_{\cal W}= x^{m}u+x^{-m}u^{-1} \ . 
\end{align}
For the minimally charged 't Hooft line operator $L={\cal  T}$, the difference operator $\hat{O}_L$ is given by \cite{Gang:2012yr} 
\begin{align}
\hat{O}_{\cal T} = H_{+} ( u, x)  \hat{p} + h_{m} (u, x  ) + H_{-} (u, x)  \hat{p}^{-1}, \nn
\end{align}
where \footnote{
The sign of the right hand side of eq~\eqref{hm} is changed from  the previous version of this paper. 
Since the 4-dimensional line operator indices obtained in \cite{Gang:2012yr} is not dependent on the sign of $h_m(u,x)$,
the change does not affect discussions in  there. 
However, since the integral equation eq~\eqref{SU(2) N_4=4 domain} is dependent on the sign of $h_m(u,x)$,
the solution eq~\eqref{4flavor-domain} is changed from our previous version. 
We regard that the current sign in eq~\eqref{hm} is more sensible
since the resultant domain wall index shows the expected symmetry under the exchange of $SU(2)$ and ${}^L SU(2)$. 
}
\begin{align}
 H_{ \pm} ( u ,  x) & =  \frac{ x  ( 1 -  u^{\mp 1} x^{ \mp m - 1 } )^4 }{ ( 1- u^{\mp2} x^{\mp 2m-2})( 1- u^{\mp 2} x^{ \mp 2m})} \ ,  \nn 
\\
h_{m} (u, x)  & =  \frac{8 x^{ m} u }{ (1+ x^{m-1} u )(1+ x^{m+1} u)  }. 
\label{hm}
\end{align}
For the duality element $\varphi=S$ which maps ${\cal W}$ to ${\cal T}$ and vice versa,  the equation \eqref{general formula} becomes
\begin{align}
&\sum_{m^\prime} \oint [du^\prime]_{m^\prime} I_{T[\Sigma_{0,4},\varphi]} (u,m;u^\prime,m^\prime) L(\hat{O}_{\cal W},\hat{O}_{\cal T}) \cdot \Pi_m (T_{0,4}) (u^\prime) \nn
\\
&=L(\hat{O}_{\cal T},\hat{O}_{\cal W})  \cdot \Pi_{m} (T_{0,4})  (u) \; \textrm{ for arbitrary polynomials  $L(x,y)$.} 
\label{SU(2) N_4=4 domain}
\end{align}
From the equations, one can determine some of $I_{T[\Sigma_{0,4},S]}$. For example, in the zero magnetic flux sector 
the index is given as 
\begin{align}
& I_{T[\Sigma_{0,4},S]}(u,m=0;u^\prime,m^\prime=0) := I_{T[ \Sigma_{0,4},S]} (u, u^{ \prime}) \nn \\
&  = 1 + 4 \left(\chi_{ \frac{1}{2}} (u)+  \chi_{ \frac{1}{2}} (u^{ \prime}) \right) x+
\left( -16+ 9 \left( \chi_1 (u) + \chi_1 (u^{ \prime }) \right) + 8 \chi_{ \frac{1}{2}} (u) \chi_{ \frac{1}{2}}(u^{ \prime}) 
\right) x^2 \nn \\
&   +\left(-36 \left( \chi_{ \frac{1}{2}}(u) + \chi_{ \frac{1}{2}} (u^{ \prime}) \right)
+ 16 \left( \chi_{ \frac{3}{2}}(u) + \chi_{ \frac{3}{2}} (u^{ \prime} ) \right)
+ 12 \left(  \chi_{1} (u) \chi_{ \frac{1}{2}}(u^{ \prime})+ \chi_{\frac{1}{2}} (u) \chi_1 (u^{ \prime}) \right)
\right) x^3 + O(x^4)
\label{4flavor-domain}
\end{align}
where $ \chi_j $ is the character of  $2j+1$ dimensional representation of $SU(2)$, $\chi_j (u) = \sum_{n=- j}^{ j} u^{2n} $. 
Here we use the fact that when $m=m^\prime=0$, the $SU(2)\times ^{L}SU(2)$ symmetry is unbroken 
and thus the index should be written as characters of two $SU(2)$s.

Let us compare the result in eq~\eqref{4flavor-domain} with the superconformal index for the 3d theory proposed  in \cite{Teschner:2012em} 
obtained by interpreting the squashed three sphere partition function of $T[\Sigma_{0,4},S]$ as a gauge theory partition function. 
The proposed  theory is 3d ${\cal N}=2$ $SU(2)$ super Yang-Mills theory with  six fundamental chiral multiplets and eight neutral chiral multiplets. 
One can read the charges of chiral fields under six Cartans of the global symmetry from the expression of the partition function  
given in  \cite{Teschner:2012em}. In table~\ref{tab:T04}, we list the charges under two Cartans of global symmetry, denoted here by $U(1)_{\rm bot} \times U(1)_{\rm top}$,  which shall correspond to two Cartans of $ SU(2) \times {}^L SU(2)$ global symmetry of $T[\Sigma_{0,4},S]$. 
\begin{table}[htbp]
   \centering
   \begin{tabular}{@{} l | cc |  c @{}} 
      \toprule
$SU(2)$     & $U(1)_{ \rm bot}$ & $U(1)_{\rm top}$ & number of chiral multiplets \\
      \midrule
      ${\bf 2}$ & $ \frac{1}{2}$ & 0 & 4   \\ 
      \cmidrule{2-4}
      & $- \frac{1}{2}$ & 1 & 1 \\ 
         \cmidrule{2-4}
      & -$ \frac{1}{2} $ & -1 & 1 \\ 
      \midrule 
      1 & -1 & 0 & 3  \\ 
      \cmidrule{2-4} 
      & 1& 0 & 1 \\
         \cmidrule{2-4} 
      & 2 & 0 & 1 \\ 
         \cmidrule{2-4} 
      & -2 & 0 & 1 \\
         \cmidrule{2-4} 
      & 0 & 1 & 1 \\
         \cmidrule{2-4} 
      & 0 & -1 & 1 \\
      \bottomrule
   \end{tabular}
   \caption{ Charges of chiral fields in a theory found to be dual to $T[ \Sigma_{0,4}]$ in  \cite{Teschner:2012em}. $SU(2)$ denotes the gauge group. 
   }
   \label{tab:T04}
\end{table}

The index for the 3d theory can be obtained using the general prescription in \cite{Imamura:2011su}. The single particle index is given as 
\begin{align}
I^{sp}_{YM} (x, u, u^{ \prime}; \zeta) = & \left( 4 \frac{u^{\frac{1}{2}} x^{\frac{1}{2}} - u^{-\frac{1}{2}} x^{\frac{3}{2}} }{1-x^2} + 
  \frac{ u^{-\frac{1}{2}}  x^{\frac{1}{2}} - u^{\frac{1}{2}}  x^{\frac{3}{2}}}{1-x^2} (u^{ \prime} + {u^{ \prime}}^{-1})
\right) 
   x^{|s|}
 (\zeta + \zeta^{-1})
\nn \\
& + 2 \frac{ x}{1 - x^2} (u^{-1} - u) \, , 
\end{align}
where the first line comes from the fundamental chiral multiplets and the second line from the two neutral chiral multiplets.
Contributions from the other six neutral chiral multiplets cancel each other. 
Here we assigned the R-charge of fundamental (resp. neutral) chiral fields to be $\frac{1}{2}$ (resp. $1$). 
$\zeta$ and $s$ are 
fugacity and magnetic flux for $SU(2)$ gauge group. The index can be obtained as, 
\begin{equation}
I_{YM} (u, u^{ \prime}) = \sum_{s=0}^{ \infty} 
 \oint [d \zeta] _s u^{- |s|} x^{ |s|} 
PE [ I^{sp}_{YM} (x, u, {u^{ \prime}} ;  \zeta) ] \, . 
\label{ind:T6}
\end{equation}
where $[d \zeta]_s$ is $SU(2)$ Haar measure given in eq~\eqref{Index multiplication}.
$x^{ |s|}$ and $u^{- |s|}$ in eq~\eqref{ind:T6} comes from zero-point contributions to the energy and flavor charge, respectively.
We check that up to several orders in $x$, the index obtained in eq~\eqref{ind:T6} coincides with the index in eq~\eqref{4flavor-domain}
\begin{equation}
I_{YM}(u, u^{ \prime})= I_{T[\Sigma_{0,4},S]} (u,u^{ \prime}) \, . 
\end{equation}
The result provides further evidence that the theory proposed in  \cite{Teschner:2012em} is dual to $T[ \Sigma_{0,4}, S]$. 
\\
\\
\\
\noindent{\bf Acknowledgements}
We are grateful to Hee-cheol Kim, Sung-soo Kim,  Hiroaki Nakajima, Sangmin Lee,  Jaemo Park, Jaewon Song, Kazuo Hosomichi, Zhao-Long Wang for helpful discussions. KL would like to thank the organizer  of Mathematical Aspects of String and M-theory workshop in Isaac Newton Institute for mathematical Sciences where this work is finished. 
This work is supported by  the National Research Foundation of Korea Grants    2006-0093850 (KL), 2009-0084601 (KL), 2010-0007512 (DG) and 2005-0049409 through the Center for Quantum Spacetime(CQUeST) of Sogang University (KL).
\appendix

\section{Index for $T[SU(2), \varphi = S^2]$}  \label{sec:I[T[SU[2],S2]]} \label{app:1}
Using the prescription in sec \ref{sec: T[SU(2)]} (see eq~\eqref{tsu2},\eqref{index for STk}), we should be able to   calculate the index for $T[SU(2), \varphi= S^2]$ theory. 
From several experiences using Mathematica, we found the followings 
\begin{align}
&I_{T[SU(2), S^2]} (u, m;u^\prime,m^\prime) = \delta_{m,m'}\mathbb{I} (m,u,u^\prime)\;, \;\textrm{where} \nn
\\
&\mathbb{I} (m=0,u,u^\prime) = \sum_n \chi_{n}(u)\chi_{n}(u^\prime) + \kappa(x)  \big{[} \chi^{odd}_{\infty} (u)\chi^{odd}_{\infty} (u^\prime) +\chi^{even}_{\infty} (u)\chi^{even}_{\infty} (u^\prime) \big{]} \;, \nn
\\
&\mathbb{I} (m>0,u,u^\prime) = \frac{x^{2|m|}}{(1-x^{2|m|} (u^\prime)^2)(1-x^{2|m|} (u^\prime)^{-2})} \sum_{n\in \mathbb{Z}} (\frac{u}{u^\prime})^n \;.
 \nn 
\end{align}
Here we define $SU(2)$ characters $\chi_n$ as 
\begin{align}
&\chi_n (u) := u^{-n}+u^{-n+2}+\ldots + u^{n-2}+u^n\;, \nn
\\
&\chi^{odd}_{\infty} :=\sum_{k =-\infty}^\infty u^{2k+1}\;,\; \chi^{even}_{\infty} :=\sum_{k =-\infty}^\infty u^{2k}\;.
\end{align}
The full analytic  expression  for $\kappa(x)$ is not determined. Listing a few lowest order in $x$, $\kappa(x)$ is given by
\begin{align}
\kappa(x)=2x+ 4x^3 - 2x^4+\ldots
\end{align}
One interesting property for $I_{T[SU(2),S^2]}$ is that it acts as the identity operator  on general $f(u,m)$, $I_{T[SU(2),S^2]}\odot f = f$,  if $f$ satisfies the following conditions 
\begin{enumerate}
\item  $f$ can be written as a Laurent series about zero, $f(u,m)= \sum_{e \in \mathbb{Z}} c_{m, e} u^e $.

 \item For $m=0$, $f$ can be decomposed into $SU(2)$ characters in $u$. Equivalently, $c_{m=0,e } = c_{m=0,-e } $.
\end{enumerate}
It is straight forward to verify this. Note that a 3d index $I_{T[SU(2),\varphi]} (u,m; u^\prime,m^\prime)$ for an arbitrary $\varphi$ satisfies  these conditions. Thus,  $I_{T[SU(2), S^2]}$ acts on the 3d index $I_{T[SU(2), \varphi]}$ as the identity operator. The second property for the 3d indices follows from the fact that the $SU(2)$ (whose Cartan is conjugate to chemical potential $u$) global symmetry is unbroken when background monopole flux $m$ is $0$.

A similar property is found for the $S^3$ partition function for $T[SU(N)]$ theory (see eq (2.24) of \cite{Nishioka:2011dq}).

\section{Index for $T[SU(3)]$} \label{app:2}

In this section, let us consider the index for $T[SU(3)]$ theory and its relation with the index for 4d ${\cal N}=4$ SYM with $G=SU(3)$. 

Let  $U={\rm diag} (U_1, U_2, U_3)$ 
and ${\bf m} ={\rm diag} (m_1,m_2,m_3)$  be the chemical potential and monopole charge for the $SU(3)$ flavor symmetry, 
imposing conditions $U_1 U_2 U_3 =1$ and $m_1 +m_2+m_3=0$. 
Also let  us use $U^{ \prime}, {\bf m}^{ \prime}$ for the chemical potential/monopole charge for the ${}^L SU(3)$ quantum symmetry. 
Then the index for $T[SU(3)]$ can be written  as follows 
\begin{align}
& I_{T[SU(3)]} ( U,{\bf m}, U^{ \prime} ,{\bf m}^{ \prime}, \eta )  \nn \\
&= \sum_{  \sigma }  \sum_{s_1 \geq s_2}
\oint^3    [d  \zeta][ d  z]   \left(\frac{U_1^{ \prime}}{ U_2^{ \prime}} \right)^{ \sigma}  \left( \frac{ U^{ \prime}_3}{U^{ \prime}_1} \right)^{s_{1}+s_{2}}
\zeta^{ m_1^{ \prime} - m_2^{ \prime} } (z_1 z_2)^{ m_3^{ \prime} - m^{ \prime}_1  } 
 x^{ \epsilon_0}  \eta^{ - \epsilon_0} \nn \\
& \times PE[   \frac{ x^{\half} \eta^{ \half} - x^{ \frac{3}{2}} \eta^{ - \half} }{1-x^2} 
\left(  \sum_{i=1}^2 x^{|\sigma-s_{i}|} ( \zeta z_{i}^{-1} + \zeta^{-1} z_{i} ) 
 +  \sum_{i=1}^2 \sum_{j=1}^3 x^{|s_{i}-m_j|} ( z_{i} (U_{j})^{-1} + z_{i}^{-1} U_{j} )  \right) ] 
 \nn \\
 & \times PE [ \frac{ x ( \eta^{-1} -  \eta )}{ 1- x^2 }(1+ 2+ x^{ | s_1 - s_2|}( z_1 z_2^{-1} + z_1^{-1} z_2) ) ]
\label{tsu3}
\end{align}
where ($\zeta$,$\sigma$) and ($\textrm{daig}(z_1, z_2)$, $\textrm{diag}(s_1, s_2)$) are (chemical potential, monopole charge) for the gauge group $U(1) $ and $ U(2)$. 
$[d \zeta]$ and $[ d z]$ are Haar measures of $U(1) $ and $U(2)$  in the monopole background, 
\begin{equation}
[d \zeta] = \frac{ d \zeta}{ 2 \pi i \zeta}, \qquad [ d z] =  \frac{1}{ (sym.)} \left( \prod_{i=1}^2 \frac{ d z_i}{ 2 \pi i z_i } \right) \prod_{i  \neq j} (1- x^{|s_i - s_j|} z_i z_j^{-1} ) . 
\end{equation}
The contour integrals are over small circles around $ \zeta=0, z_i =0$. 
The symmetric factor $(sym.)$ is $1$ if $ s_1 \neq s_2$, and $2$ if $s_1 = s_2$. $ \epsilon_0$ is the Casimir energy given as 
\begin{align}
\epsilon_0 &= 
\half \sum_{i=1,2} | \sigma - s_i |
+ \half \sum_{i=1,2} \sum_{j=1,2,3} | s_i - m_j|   
- |s_1 - s_2 |  \, , 
\nn
\end{align}
where the last term comes from the vector multiplet of $U(2)$, and the other terms are from hypermultiplets. 
The summations are over monopole charges satisfying Dirac quantization, $ \sigma - s_i \in {\mathbb Z}$ and $s_i - m_j \in {\mathbb Z}$. For instance, if ${\bf m}= {\rm diag}(\frac{2}{3}, - \frac{1}{3}, - \frac{1}{3})$, the sum of $\sigma$ and $s_i$ run over $\frac{2}{3} + {\mathbb Z}$. We impose a condition $s_1 \geq s_2$ on the summation using Weyl symmetry of $U(2)$ gauge group. 

Expanding in $x$, the first two terms for ${\bf m}={\bf m}^{ \prime}=0$ are 
\begin{align} 
I_{T[SU(3)]} (U, {\bf m}, U^{ \prime},{\bf m}^{ \prime} )|_{{\bf m}={\bf m}^{ \prime}=0} &=
1+( \chi_{adj} ( U) + \chi_{adj} (U^{ \prime})) x  + O(x^2) 
\label{x1}
\end{align}
where $ \chi_{adj}(U)= (U_1 + U_2)(U_1 + U_3)(U_2+U_3) |_{U_1 U_2 U_3=1}$ denotes the character of the adjoint representation of $SU(3)$. 
In eq~\eqref{x1}, the index is invariant under the exchange of $U$ and $U^{ \prime}$ to the displayed order in $x$. The invariance is indeed checked to several orders in $x$ using mathematica. 

Since the 4d half-index  \eqref{half-index} for $G=SU(3)$ and 3d $T[SU(3)]$ index \eqref{tsu3} are both explicitly given, we can check the condition eq~\eqref{condition}. For $ \Pi(U) := \Pi_{{\bf m}=0} (U)$, we check the following holds up to several orders in $x$ 
\begin{align}
\Pi (U) &= \oint [dU^{ \prime}] \  I_{T[SU(3)]} (U, {\bf m}, U^{ \prime},{\bf m}^{ \prime}, \eta )|_{{\bf m}={\bf m}^{ \prime}=0, \eta=1} \ \Pi (U^{ \prime}) \ .  \nn 
\end{align}

Let us now consider how S-duality acts on the fundamental Wilson line in 4d ${\cal N}=4$ $SU(3)$ theory. 
Since the fundamental representation of $SU(3)$ is complex,
an insertion of the fundamental Wilson line is distinguishable from that of the anti-fundamental Wilson line
unlike to $SU(2)$ gauge group.  
The 't Hooft line operator $\hat{O}_{(1,0)}$ (resp. $\hat{O}_{(1,1)}$) with magnetic charge ${\rm diag} (\frac{2}{3}, - \frac{1}{3}, - \frac{1}{3}) $ (resp. ${\rm diag}(\frac{1}{3}, \frac{1}{3}, - \frac{2}{3})$)
is obtained in \cite{Gang:2012yr}.  In several orders in $x$, we find that the followings hold 
\begin{equation}
\begin{cases}
& \sum_{{\bf m}^{ \prime}} \oint [ dU^{ \prime}] 
 I_{T[SU(3)]} (U, {\bf m}, U^{ \prime} , {\bf m}^{ \prime} )\ 
\chi_{\bf 3} (U^{ \prime}) \Pi_{{\bf m}^{ \prime}} (U^{ \prime}) = \hat{O}_{(1,0)}\cdot \Pi_{{\bf m}} (U),  \\
&  \sum_{{\bf m}^{ \prime}} \oint [ dU^{ \prime}] 
I_{T[SU(3)]} (U, {\bf m}, U^{ \prime} , {\bf m}^{ \prime} ) \ 
\chi_{\bar{ \bf 3}} (U^{ \prime}) \Pi_{{\bf m}^{ \prime}} (U^{ \prime}) = \hat{O}_{(1,1)} \cdot \Pi_{{\bf m}} (U) \, , 
\end{cases}
\label{su3-line}
\end{equation}
where $ \chi_{\bf 3}(U)= U_1 + U_2 + U_3 |_{U_1 U_2 U_3 =1} $ and 
$ \chi_{\bar{3}}(U)=U_1^{-1} + U_2^{-1}+U_3^{-1} |_{U_1 U_2 U_3=1}$ are 
characters for ${\bf 3}$ and $\bar{\bf 3}$ representations of $SU(3)$. The first equation of eq~\eqref{su3-line} implies that S-duality maps the fundamental Wilson line, ${\cal W}_{\bf 3}$,
to the  't Hooft line with magnetic charge ${\rm diag} ( \frac{2}{3}, - \frac{1}{3}, - \frac{1}{3})$, ${\cal T}_{\rm diag(\frac{2}{3}, - \frac{1}{3} , - \frac{1}{3} )}$. 
The second equation implies that S-duality 
maps the anti-fundamental Wilson line, ${\cal W}_{\bar{\bf 3}}$, to the 't Hooft line with charge ${\rm diag} ( \frac{1}{3}, \frac{1}{3}, - \frac{2}{3})$,
${\cal T}_{{\rm diag}( \frac{1}{3}, \frac{1}{3}, - \frac{2}{3})}$. 
Also, we can check that the followings hold in several orders in $x$, 
\begin{align} 
\begin{cases}
& I_{T[SU(3)]} \odot ( \hat{O}_{(1,0)} \cdot \Pi )_{{\bf m}} (U) = \chi_{\bar{\bf 3}} (U) \Pi_{\bf m} (U),  \\
& I_{T[SU(3)]} \odot ( \hat{O}_{(1,1)} \cdot \Pi)_{{\bf m}} (U)  = \chi_{\bf 3} (U) \Pi_{\bf m} (U)  \, . 
\end{cases}
\label{su3-line-2}
\end{align}
The equations imply that S-duality maps ${\cal T}_{{\rm diag}( \frac{2}{3}, - \frac{1}{3}, - \frac{1}{3})}$  to ${\cal W}_{\bar{\bf 3}}$, 
and ${\cal T}_{{\rm diag}( \frac{1}{3}, \frac{1}{3} , - \frac{2}{3} )}$ to ${\cal W}_{\bf 3}$. 

The action of S-duality on line operators can be summarised as follows 
\begin{eqnarray}
{\cal W}_{{\mathbf 3}} & \to & {\cal T}_{{\rm diag}( \frac{2}{3}, - \frac{1}{3}, - \frac{1}{3})}  \nonumber \\
\uparrow & & \downarrow \nonumber \\
{\cal T}_{{\rm diag}( \frac{1}{3}, \frac{1}{3}, - \frac{1}{2})}  &  \leftarrow & {\cal W}_{\bar{\mathbf 3}} \nonumber
\end{eqnarray}
which is consistent with $S^4 = 1$. 
 It would be interesting to check this S-duality map of line operators in other gauge groups.

\providecommand{\href}[2]{#2}\begingroup\raggedright\endgroup

\end{document}